\documentclass[journal]{IEEEtran}
\usepackage[normalem]{ulem}
\usepackage{amsmath}
\usepackage{color}
\usepackage{array}
\usepackage{stfloats}
\usepackage{amsthm} 
\usepackage{amssymb}
\usepackage{float}
\usepackage{nccmath}
\usepackage{graphicx}
\usepackage[linesnumbered,ruled]{algorithm2e}
\usepackage{setspace}
\usepackage{booktabs}
\usepackage{mwe}
\usepackage{cite}
\usepackage{amsmath,amssymb,amsfonts}
\usepackage{graphicx}
\usepackage{textcomp}
\usepackage{xcolor}
\usepackage{physics}
\usepackage{bm}
\usepackage[utf8]{inputenc} 

\usepackage{algorithmicx}
\usepackage{xspace}
\usepackage{subfigure}
\def\BibTeX{{\rm B\kern-.05em{\sc i\kern-.025em b}\kern-.08em
    T\kern-.1667em\lower.7ex\hbox{E}\kern-.125emX}}
\begin{document}
\newcommand{\BfPara}[1]{{\noindent\bf#1.}\xspace}
\newcommand\mycaption[2]{\caption{#1\newline\small#2}}
\newcommand\mycap[3]{\caption{#1\newline\small#2\newline\small#3}}

\renewcommand{\baselinestretch}{.88}

\title{Attention-based Reinforcement Learning for Real-Time UAV Semantic Communication}

\author{Won Joon Yun, Byungju Lim, Soyi Jung, Young-Chai Ko, $^\dagger$Jihong Park, Joongheon Kim, and $^\ddagger$Mehdi Bennis
\thanks{W. Yun, B. Lim, S. Jung, Y.-C. Ko, and J. Kim are with the School of Electrical Engineering, Korea University, Seoul 02841, Korea (email: \{ywjoon95\;,limbj93\;,jungsoyi\;,koyc\;,joongheon\}@korea.ac.kr).}%
\thanks{$^\dagger$J. Park is with the School of Information Technology, Deakin University, Geelong, VIC 3220, Australia (email: jihong.park@deakin.edu.au).}%
\thanks{$^\ddagger$M. Bennis is with the Centre for Wireless Communications, University of Oulu, Oulu 90014, Finland (email: mehdi.bennis@oulu.fi).}%
\thanks{J. Park and J. Kim are the corresponding authors of this paper. }%
\thanks{This research was supported by Institute for Information \& Communications Technology Promotion (IITP) grant funded by the Korea government (MSIT) (No.2018-0-00170, Virtual Presence in Moving Objects through 5G).}%
}

\maketitle

\begin{abstract}
In this article, we study the problem of air-to-ground ultra-reliable and low-latency communication (URLLC) for a moving ground user. This is done by controlling multiple unmanned aerial vehicles (UAVs) in real time while avoiding inter-UAV collisions. To this end, we propose a novel multi-agent deep reinforcement learning (MADRL) framework, coined a graph attention exchange network (GAXNet). In GAXNet, each UAV constructs an attention graph locally measuring the level of attention to its neighboring UAVs, while exchanging the attention weights with other UAVs so as to reduce the attention mismatch between them. Simulation results corroborates that GAXNet achieves up to 4.5x higher rewards during training. At execution, without incurring inter-UAV collisions, GAXNet achieves 6.5x lower latency with the target 0.0000001 error rate, compared to a state-of-the-art baseline framework.
\end{abstract}
\section{Introduction}
Unmanned aerial vehicles (UAVs) can provision agile and mobile network infrastructure \cite{TVT2021_Orchestration,KimSPAWC:18}, and enable ultra-reliable and low-latency communication (URLLC) even under time-varying environments and tasks, such as moving target sites in disaster scenes and transformable assembly lines in smart factories, to mention a few. As opposed to the stationary and fixed nature of 5G URLLC, such non-terrestrial URLLC systems in beyond 5G are time-varying and physically reconfigured, mandating to co-design control and communication in real time \cite{park2020extreme}. 

To this end, machine intelligence is envisaged to play a crucial role. Without any central coordination, intelligent agents can promptly react to local environmental changes, thereby reducing latency while saving radio resources \cite{park2018wireless}. To imbue the intelligence into multiple interactive agents, multi-agent deep reinforcement learning (MADRL) is a promising solution  \cite{NIPS2017_MADDPG,AAAI2020_G2ANet}, wherein each agent runs deep learning so as to maximize its expected long-term reward by carrying out actions for its given state observations, i.e., decision-makings on state-action policies. Depending on how to train and execute the deep learning models, MADRL is broadly categorized into three types as elaborated next. 

First, \textit{centralized MADRL} is the case wherein all agents send their observations to a central entity to build a global policy determining the actions of the entire agents. MADDPG~\cite{NIPS2017_MADDPG}, CommNet~\cite{NIPS2016_CommNet}, and G2ANet~\cite{AAAI2020_G2ANet} fall in this category. These algorithms presumably achieve the highest reward, albeit at the cost of frequently exchanging a non-negligible amount of information on the states, actions, and rewards, which are thus far from meeting stringent latency constraints in URLLC. Second, \textit{fully decentralized MADRL} is the type in which every agent trains and executes its policy without exchanging any information with other entities. These include IQL~\cite{ICML1993_IQL} and I-DQN~\cite{PLOS2017_IDQN}, which however may not guarantee reliability due to the lack of understanding the interactions among agents. Lastly, \textit{centralized training and decentralized execution (CTDE)} MARL lies between the aforementioned two extremes. Following the actor-critic architecture~\cite{NIPS00_A2C}, an actor model stored at each agent determines its policy for both training and execution, while a centralized critic model evaluates the reward of all agents only during training.

In this work we aspire to build a novel CTDE MADRL framework for UAV aided beyond-5G URLLC, as visualized in Fig.~1. To this end, we first delve into the opportunities and limitations of the existing CTDE MARL frameworks. In Differentiable Inter-Agent Learning (DIAL) \cite{NIPS2016_c7635bfd}, while taking actions, the agents exchange clean-slate messages passed through their actor models. During training, these messages are progressively turned into meaningful representations for better inter-agent collaboration, hereafter referred to as \emph{semantic representations (SRs)}, which is an analog of children's developing language-based cues as they grow. These emergent SRs are effective in achieving higher rewards at execution, yet its starting from clean-slate messages is too inefficient to outperform other state-of-the-art CTDE frameworks. Alternatively, during training, the agents in QMIX \cite{ICML2018_QMIX} and CSGA \cite{SMC2021_CSGA} exchange global states and local graph attention, respectively, thereby achieving competitive performance. However, they do not share any local information during execution, so cannot reach the full potential of CTDE MADRL. 

By integrating DIAL's emergent SR learning into graph attentive CSGA, in this work we propose a novel CTDE MADRL framework, termed a \emph{graph attention exchange netowrk (GAXNet)}. In CSGA, each agent $n$ at every time $t$ locally constructs a star-topological graph as shown in Fig.~2a. The edge weight $w_{n,m}^t$ increases with the level of attention to the agent $m$ when the agent $n$ takes its action. For mutually symmetric agent interactions (e.g., collision, repulsion, etc.), the rationale being what I attend to you should ideally be the same as what you attend to me, i.e., $w_{n,m}^t=w_{m,n}^t$. This condition is often violated in CSGA due to its local attention graph construction. Inspired by DIAL, we overcome such inter-agent attention mismatch via developing and exchanging emergent SRs.

To be precise, the agent~$n$ runs an SR encoder whose input is the agent $n$'s current attention to the agent $m$ and its counterpart attention to the agent $n$ from the agent $m$ in the previous time slot, resulting in the output SR $\bar{w}_{n,m}^t$, i.e., $\bar{w}_{n,m}^t \overset{\text{SR}}{\leftarrow} \{w_{n,m}^t,\bar{w}_{n,m}^{t-1}\}$ as illustrated in Fig.~2b. This SR $\bar{w}_{n,m}^t$ is exchanged across agents for constructing SR $\bar{w}_{n,m}^{t+1}$ in the next time slot. In contrast to DIAL that initially exchanges meaningless messages, GAXNet exchanges semantically meaningful attention weights from the beginning, promoting better SR emergence. Consequently, the agent $n$ determines its action at time $t$ based on the emergent SR $\bar{w}_{n,m}^t$, as opposed to $w_{n,m}^t$ in CSGA, thereby reflecting the semantic importance of the attention at both agent sides, aiming at ensuring $\hat{w}_{n,m}^t\approx \hat{w}_{m,n}^t$ for all interacting agents. 

To show the effectiveness of GAXNet in UAV aided beyond-5G URLLC, we study a scenario where a moving ground URLLC user is supported at least by one of the multiple fixed-wing UAVs. These UAVs aim to hover within a range guaranteeing URLLC requirements, referred to as \emph{URLLC range}, while avoiding inter-UAV collisions. Simulations validate that compared to QMIX \cite{ICML2018_QMIX}, the proposed GAXNet significantly reduces inter-UAV collision occurrences, and achieves by up to $4.5$x higher reward for $5,000$ epochs during training, where the reward is increased when the agent satisfies the latency and reliability requirements, and is penalized when an inter-UAV collision occurs. At execution, GAXNet achieves up to $6.5$x lower latency with the target $10^{-7}$ error rate, compared to QMIX that fails to guarantee the target error rate. 

\begin{figure*}[t!]
\centering\label{fig:abstract}
    \includegraphics[width=1\linewidth]{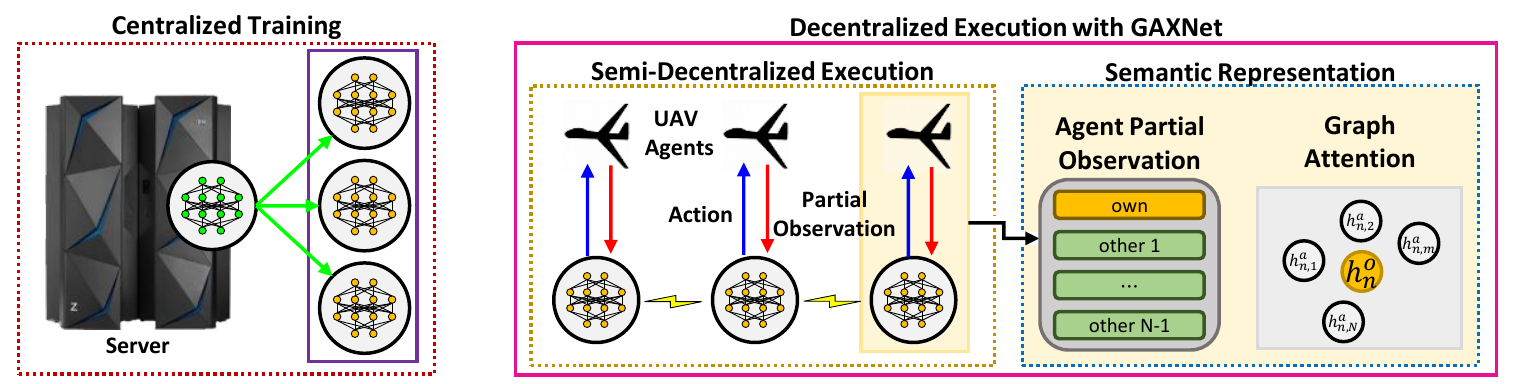} 
    \caption{A schematic illustration of the proposed graph attention exchange network (GAXNet) and its application to a multi-UAV emergency network.}
    \vspace{-5pt}
\end{figure*}

\section{Preliminaries: MADRL and Self Attention}\label{sec:2}

GAXNet relies on two key principles, CTDE MADRL and the self-attention mechanism, which are described in the following subsections.

\begin{figure}[t!]
    
    \subfigure[Local attention graph construction at the agent $n$.]{\includegraphics[width=1\columnwidth]{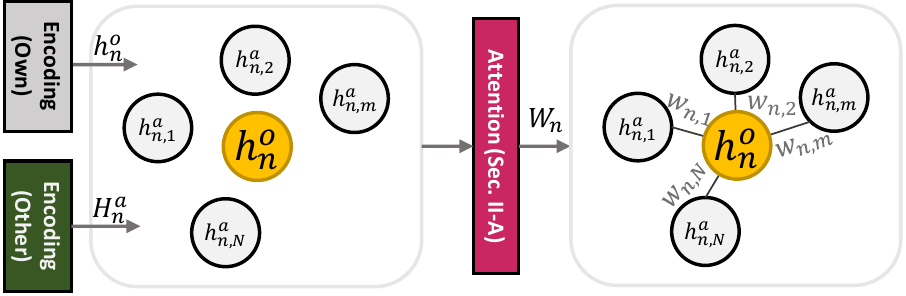}}\label{fig:semanticrepresentation}
    
    \subfigure[SR encoding at the agents $n$ and $m$.]{\includegraphics[width=1\columnwidth]{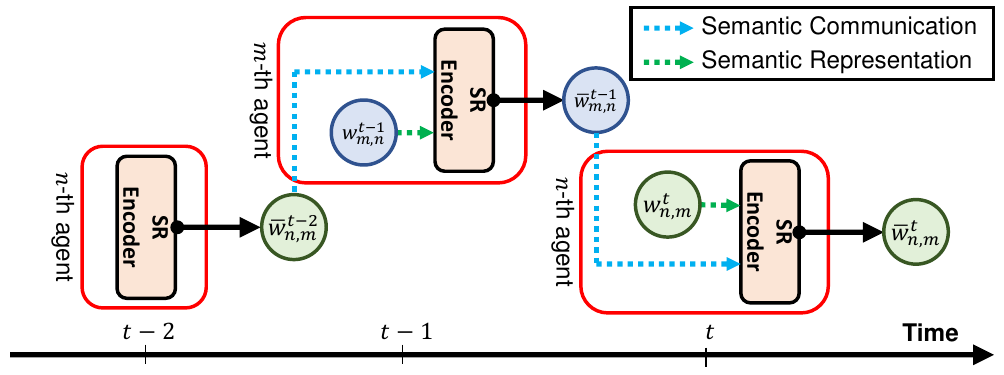}} \label{fig:semanticcommunication}
    \vspace{-5pt}
    \caption{The operations of GAXNet: (a) local attention graph construction at the agent $n$ and (b) semantic representation (SR) encoding at the agents $n$ and $m$ for $3$ consecutive time slots.}
    \vspace{-5pt}
\end{figure}
\subsection{Self-Attention}\label{sec:2-1}
Let's denote the input as $\mathcal{X} = \left\{\bm{x}_1, \cdots, \bm{x}_k, \cdots, \bm{x}_K \right\}$. The input goes through the encoding process that converts into a trainable form. The process appears in the form of a linear combination, and the formula is summarized:
\begin{equation}
\mathcal{H}^{enc} = \bm{W}^{enc}\cdot\mathcal{X} + \bm{b}^{enc},
\label{eq:enc1}
\end{equation}
where $\bm{W}^{enc}$ and $\bm{b}^{enc}$ stand for the weight and bias of the encoding layer, respectively.
Self-attention is the process of determining the coefficient between inputs using scale-dot production~\cite{attention_seq2seq}. It consists of query layer, key layer and, value layer. Note that query layer, key layer and, value layer are denoted as $l^q$, $l^k$, $l^v$.
Self-attention converts the received hidden variable into query, key, and value. This process can be expressed as follows:
\begin{eqnarray}
\label{eq:query}
    \textbf{Q}= l^q(\mathcal{H}^{enc}) = \left\{\bm{q}_1, \cdots, \bm{q}_k, \cdots, \bm{q}_K \right\}, \\
\label{eq:key}
    \textbf{K}= l^k(\mathcal{H}^{enc}) = \left\{\bm{k}_1, \cdots, \bm{k}_k, \cdots, \bm{k}_K \right\}, \\
\label{eq:value}
    \textbf{V}= l^v(\mathcal{H}^{enc}) = \left\{\bm{v}_1, \cdots, \bm{v}_k, \cdots, \bm{v}_K \right\},
\end{eqnarray}
where $\bm{q}_k, \bm{k}_k$, and $\bm{v}_k$ stand for transformer of query, key, and value of the input $\bm{x}_k$. By taking dot product of query $\bm{q}_k$ and other input keys $\bm{K}_{-k}\triangleq \cup_{k'\neq k}^K \{\bm{k}_{k'}\}$, the key component of other input information is obtained.
After that, we multiply by the $k$-th 's value, and the weighted message matrix is obtained. 
Formally, $W$ is given as:
\begin{equation}
    W=\textsf{Attention}(\textbf{Q},\textbf{K}^{-1},\textbf{V})
=\dfrac{\textbf{Q} \cdot \textbf{K}^{-1} \cdot \textbf{V}}{\sqrt{K}},
\label{eq:softattention}
\end{equation}
where $W$ is $K$-by-$(K-1)$ matrix which represents the weights of adjacency matrix, $\sqrt{K}$ is a scaling factor that prevents the attention value diverge and note that $\textbf{K}^{-}\triangleq \left\{\bm{K}_{-1}, \cdots, \bm{K}_{-n}, \cdots, \bm{K}_{-N} \right\}$.

\subsection{Centralized Training and Decentralized Execution MADRL}\label{sec:2-2}
In this subsection, we introduces CTDE MADRL in briefly. CTDE MADRL is mathematically modeled with Dec-POMDP. Dec-POMDP is a stochastic decision making model that is to consider the uncertainty between agents~\cite{Springer2016_POMDP}. Dec-POMDP is defined as a tuple $G=\langle S, A, P, r, Z, O, n, \gamma\rangle$. In environment, the finite set of agents is denoted as $U=\left\{u^1, u^2, \cdots, u^N\right\}$, where $N\in\mathbb{N}^1$, respectively. The global state of the environment is defined as $s\in S$. The action of $n$-th agent is defined as $a^n \in A$. Since all agents can observe local information $z\in Z$, the observation function is defined as $O(s,u): S \times U \rightarrow Z$. 

The policy of the agent $\pi^n$ takes the observation-action history $\tau^n\in T$ and decides action as $\pi^n(a^n|\tau^n): T \times A$.
In addition, the joint action of all agents is $\bm{a} \in \bm{A}$. The state transition function is denoted as $P(s',s,\bm{a}): S \times A \times S \rightarrow[0,1]$. All agents obtain same reward $r(s,\bm{a}): S \times \bm{A} \rightarrow \mathbb{R}$. The joint action value function at time $t$ is defined as follows:
\begin{equation}
    Q^{\pi}(s_t,\bm{a}_t)=\mathbb{E}\left[\sum^{\infty}_{i=t+1}\nolimits\gamma^{i-t-1}r_{i}\mid s_{t},\bm{a}_{t}\right],
\end{equation}
where $\pi$ and $\gamma \in [0, 1)$ stand for joint policy and discount factor, respectively.  

The objective of MADRL system is to maximize the cumulative reward. In order to maximize cumulative reward under the consideration of centralized training and decentralized execution and according to \cite{ICML2018_QMIX}, the joint action value function is decomposed by agent's individual utility function. The optimal joint action value function is as follows:
\begin{equation}
     \arg\max_{\bm{a}} Q^{\pi}({\bm{\tau}}  ,\bm{a}, s;\theta) =[\cdots,\arg\max_{a^n} Q^n(\tau^n,a^n;\theta^n),\cdots],
\end{equation}
where $\theta$ and $\theta^n$ are the parameters of $\pi$ and $\pi^n$, respectively. 
To summarize CTDE, the joint action value function leads all agents' optimal utility function, all agents determine their actions via their policies.

\section{UAV Automation Framework for URLLC}\label{sec:3}
\subsection{Network and Channel Model}\label{sec:3-1}
As illustrated in Fig.~1, the network under study consists of a set $\mathcal{U} \triangleq \{ u_1, \cdots, u_n, \cdots, u_N \}$ of $N$ fixed-wing UAV agents and a moving target location $c$. For simplicity, we hereafter focus only on UAV-to-ground channels, while assuming the inter-UAV communications are always successful with negligible delays. This is not a strong assumption in our experimental settings with $N=4$ (see Sec. IV) wherein the payload size to be received during a unit time slot is only $576$ bits. 

According to \cite{WCSP_2018_SNR}, the air-to-ground mean path loss for $xy$-direction distance $d$ and $z$-direction distance $h$, is given as: \begin{align}\label{PL}
    \text{PL}(d,h) &= \frac{\eta_{LoS}-\eta_{NLoS}}{1+\alpha \cdot\textsf{exp}\left(-\beta(\frac{180}{\pi}\tan^{-1}\left(\frac{h}{d}\right)-\alpha)\right)} \\ \nonumber
    &+ 10\text{log}_{10}(h^2+d^2) + 20*\text{log}_{10}\left(\frac{4\pi f_c}{c}\right)+\eta_{NLoS},
\end{align}
where $\eta_{LoS}$ and $\eta_{NLoS}$ stand for additional pathloss of LoS and NLoS. Also, $\alpha$ and $\beta$ represent environmental constant, respectively.
Then, Signal-to-noise ratio (SNR) can be calculated as follows:
\begin{equation}\label{SNR}
    \text{SNR}(d,h)=\frac{P_t}{P_n}*10^{-{\text{PL}/10}},
\end{equation} where $P_t$, $P_n$, and $W$ stand for transmit power, noise power, and bandwidth of A2G, respectively. According to \cite{PIEEE_2021_PER}, the error rate is obtained as follows:
\begin{align}\label{error_rate}
    \varepsilon = \frac{1}{\sqrt{2\pi}}\int^\infty_{f(\text{SNR})}e^{-\frac{t^2}{2}}\text{d}t=Q_f(f(\text{SNR})),
\end{align}
where $Q_f(\cdot)$ is the Q-function and $f(\text{SNR})$ is given by $f(\text{SNR})=\sqrt{\frac{W\cdot T_{\max}}{1-(1+\text{SNR})^{-2}}}\left(\ln(1+\text{SNR})-\frac{R_s\ln2}{W}\right)$, 
and $T_{\max}={L_B/W}$ and $R_s$ represent the maximum transmission time and achievable throughput, respectively.
The achievable throughput $R_s$ can be calculated as follows:
\begin{equation}
    R_s  \simeq \text{ln}(1+\text{SNR})-\frac{1-(1+\text{SNR})^{-2}}{L_B}Q_f(\varepsilon_0),\label{eq:datarate}
\end{equation} 
where $\varepsilon_0$ is the required reliability.
 
\subsection{MADRL Model}\label{sec:3-2}
Next, we introduce the multi-agent reinforcement learning  formulation (\textit{e.g.}, state, action, and reward).
\subsubsection{State space}
The state space of our proposed UAV automation framework of UAVs/target location information, and relative position information with other UAVs/targets. The position of $u_n$ at time $t$ is defined as $l^n_t$,  where $\forall l_n = (x^n_t, y^n_t), n \in [1, N]$. In addition, $l^{n,c}_t=(x^{n,c}_t, y^{n,c}_t)$ and $d^{n,c}_t$ denote the relative positions and distance for the $u_n$ with $c$, respectively. The relative position and distance of $u_n$ and $u_m$, $m\in [1,N]\backslash n $ is denoted as $l^{n,m}_t$ and $d^{n,m}_t$. In addition, when $u_n$ is observable of $u_m$, $e^{n,m}_t = 1$, and $e^{n,m}_t = 0$ when unobservable. The observation of $u_n$ at $t$ consists of position information, relative position information and distance information about the target and other agents as follows:
 
\begin{align}
    o^n_t=\underbrace{\bigcup^{1}_{i=0}\{l^n_{t-i},l^{n,c}_{t-i}, d^{n,c}_{t-i}\}}_{\text{own}}\cup\underbrace{\bigcup^N_{m\neq n}\{l^{n,m}_t,d^{n,m}_t,e^{n,m}_t\}}_{\text{other agents}}, \nonumber\label{eq:observation}
\end{align}
Note that when $u_n$ is not observable (\textit{i.e.}, $e^{n,m}_t=0$), $l^{n,m}_t,d^{n,m}_t,e^{n,m}_t$ are zero.
The true state information $s_t$ in the environment represents the observation information of all agents and the absolute position information of the target in time $t$ and $t-1$.
\begin{equation}
    s_t = \bigcup^{1}_{i=0}\left\{\bigcup^N_{n=1}\left\{o^n_{t-i}\right\}\cup l^c_{t-i}\right\}, \label{eq:state}
\end{equation}
where $l^c_{t}$ stands for the absolute position of the target.

\subsubsection{Action space}
The action space of the UAV automation framework consists of $8$ discrete actions, which is defined as $A \triangleq \{(x+\nu, y),(x-\nu, y),(x, y+\nu),(x,y-\nu),(x+\frac{1}{\sqrt{2}}\nu, y+\frac{1}{\sqrt{2}}\nu),(x-\nu, y+\frac{1}{\sqrt{2}}\nu),(x+\frac{1}{\sqrt{2}}\nu, y-\frac{1}{\sqrt{2}}\nu),(x-\frac{1}{\sqrt{2}}\nu,y-\frac{1}{\sqrt{2}}\nu)\}$, where $\nu$ is the speed of UAV. At time $t$, the action of $u_n$ is denoted as $a^n_t \in A$, where the joint action is $\bm{a_t} = \{ a^1_t , \cdots , a^n_t, \cdots , a^N_t\}$.

\subsubsection{Reward space}
 
The reward space is designed to make UAV agent quickly reach the target area and prevent collision.
For $u^n$ to guarantee the URLLC requirements, $u^n$ should reach to the target quickly, the URLLC reliability reward is
 
\begin{align} \label{eq:rwd_target}
r^{n,c}_t=
    \begin{cases}
        1, ~~~&\text{if. } d^{n,c}_t < d^c,\\
        0.05, &\text{if. } d^c \leq d^{n,c}_t < d^{n,c}_{t-1}\\
        -0.01, &\textrm{otherwise}.\\
    \end{cases}
\end{align}
where $d^c$ represents the URLLC range for URLLC reliability, respectively. If $d^{n,c}_t$ is less than $d^c$, the error rate of $u^n$ is less than the required error rate for URLLC and vice versa. In addition, the collision reward for minimizing collision between agents is presented as follows:
\begin{align}\label{eq:rwd_collision} r^{i,n,m}_t=
    \begin{cases}
        -0.5, ~~~&\text{if. }d^{n,m}_t < d^{i},\\
        0,&\textrm{otherwise}.\\
    \end{cases}
\end{align}
When the distance between $u_n$ and $u_m$ is less than $d^{i}$, collision occurs and vice versa. 
 
According to \eqref{eq:rwd_target} and \eqref{eq:rwd_collision}, the final reward at time $t$ is as follows:
 
\begin{equation}
    r_t =  \underbrace{\sum^N_{n=1}\nolimits{r^{n,c}_t}}_{\text{UAV-to-ground URLLC}} \!\!\!+\;\;\; \underbrace{\sum^N_{n=1}\nolimits\sum^N_{m\neq n}\nolimits{r^{i,n,m}_t}}_{\text{Inter-UAV collision}} 
\end{equation}

Note that all agents have a common reward.

\subsection{Policy Updates Using Graph Attention and SR Encoding}\label{sec:3-3}
In this section, we introduce how to configure policies to generate and transfer semantic messages. Since the agent's observation consists of its own unique information and partial information of other agents in Dec-POMDP, the agent information can be operated by categorization of the agent's observation into the agent's unique information and the other agent's information. 
 
$o^n_t$ is encoded by taking input in the form taken by the flatten on the out characteristic encoding layer, while partial information about other agents is stacked to encode in the other agents encoding layer as follows:
\begin{align}
    h^o_n&=f^{enc,n}_{own}(\mathsf{Flatten}(o_t)), \label{eq:enc_own}\\
    H^{a}_n&=f^{enc,n}_{oth}(\mathsf{Stack}(o^{oth,n,m})), \forall m \in [1,N]\backslash n,\label{eq:enc_oth}\\
    \text{where, }&~~o^{oth,n,m}_t\triangleq\{l^n_{t},l^{n,d}_{t}, l^{n,m}_t,d^{n,c}_{t},d^{n,m}_t,e^{n,m}_t\}, \nonumber
\end{align}
and $f^{enc,n}_{own}$, $f^{enc,n}_{oth}$ and $o^{oth,n,m}_t$ stand for $n$-th agent's own characteristics encoding layer, other agent's characteristics encoding layer, and observation about $u_m$ of $u_n$, respectively. In addition, $H^a_n$ is a set of vector, which represents to $H^a_n=\cup^{N}_{m\neq n}\{h^{a}_m,\}$. $h^o_n$, and $h^a_m$ have a vector size of $1\times J_1$, and $H^{a}_n$ has a size of $N-1 \times J_1$, where $J_1$ stands for the output dimension of encoding layer.

Fig.~2(a) illustrates the method that utilizes self-attention  using $h^o_n$ and $H^a_n$ which are provided via \eqref{eq:enc_own}--\eqref{eq:enc_oth}.
For $\mathcal{H}^{enc}$ in Sec.~\ref{sec:2-1}, $\mathcal{H}^{enc}$ is redefined as $\mathcal{H}^{enc}=\{h^a_1, \cdots, h^a_{n-1}, h^o_n,h^a_{n+1},\cdots,h^a_N\}$.
In \eqref{eq:query}--\eqref{eq:value}, $l^q$, $l^k$, $l^v$ are replaced by the query layer $l^q_n$, key layer $l^k_n$, and value layer $l^v_n$ of the $u_n$. Therefore the query, key, and value can be obtained through $q_n = l^q_n(h^o_n)$, $k_n = l^k_n(H^a_n)$, $v_n = l^v_n(H^a_n)$.
The size of each query, key, and value represents $1\times J_2$, $N-1\times J_2$, and $N-1\times J_2$. $J_2$ represents the attention dimension, respectively.
The weight of semantic representation for other agents can be obtained by taking \eqref{eq:softattention}, which is denoted as $W_n=\{w_{n,1},\cdots w_{n,n-1},w_{n,n+1},\cdots,w_{n,N}\}$.

Let's assume the exchange of the semantic weights $w_{n,m}$ and $w_{m,n}$ of two agents $u_n$ and $u_m$ for semantic communication.
Fig.~2(b) shows the overall process of semantic communication between $u_n$ and $u_m$.

The weight information $w_{n,m}$ which is generated by $u_n$, is the semantic weight created from the complete information of the $u_n$ and the partial information of the $u_m$ and vice versa. 
At time $t$, $u_n$ receives the $w^{t-1}_{m,n}$ which is created by the $u_m$ in the previous step $t-1$, through the semantic channel. The synthesized weight information is created by aggregating $w_{n,m}$, and $w_{m,n}$ through the SR encoder. The synthesized weight is denoted as $\bar{w}^t_{n,m}$. The process of semantic communication is as follows: 
\begin{equation}
    \bar{w}^{t}_{n,m} = 
    \begin{cases}
        \mathsf{RNN}^n(w^t_{n,m}, \bar{w}^{t-1}_{m,n}), & \text{if.}~ e^{n,m}_t = 1\\
        \mathsf{RNN}^n(w^t_{n,m}, \vec{0}), &~~~\textit{otherwise.}\\
    \end{cases}
\end{equation}
where $\mathsf{RNN}^n$ and $\vec{0}$ stand for SR encoder, and $N-1 \times 1$ size zero vector, respectively. \texttt{GRUCell} is used for SR encoder~\cite{GRU}. Note that $u_n$ obtains the semantic weight $w_{m,n}$ from all observable agents excluding itself (\textit{i.e.}, $e^{n,m}_t=1$). Finally, $\pi^n$ determines $Q^n(\tau^n_t,a^n_t;\theta^n)$ which indicates the utility function of $u_n$, utilizing $h^{o,n}_t$ and $\bar{W}^n_t$. 

\subsection{Centralized Training and Decentralized Execution}\label{sec:3-4}
In this subsection, we introduce a CTDE method for UAV aided URLLC systems, consisting of $N$ decentralized actors and one centralized critic denoted as $\Phi(\theta)$. QMIX architecture is adopted as the centralized critic~\cite{ICML2018_QMIX}. 
The utility function of all agents, and environment true state are the input of $\Phi(\theta)$.  
In addition, the observation and action of all agents, current state, reward, the next observation of all agents, and the next state are stored as tuple $b =\langle O,S,\bm{A},R,O',S'\rangle$ to the replay buffer. 
In training phase, $B=\{b_1,\cdots,b_i,\cdots,b_I\}$ are sampled from the replay buffer. $B$ is training data to train the parameters of actors and critic. Temporal difference error is used as loss function~\cite{Arxiv2013_Atari}, i.e., 
\begin{equation}
    L(\theta)= \sum_{b\in B}\left[r+ \gamma \max_{\bm{a}}Q^{\Psi}(\bm{\tau}',\bm{a}',s';\theta^{-})-Q^{\Phi}(\bm{\tau},\bm{a},s;\theta)\right]^2,
\end{equation}
where $\Phi(\theta^{-})$ stands for target network. 
The agent is trained with the direction to maximize the joint-action value function, and if the agent can communicate with others, it can reinforce cooperation through semantic communications. 
\section{Performance Evaluation}\label{sec:4}

\begin{table}[t!]\label{tab:parameters}
\caption{The parameters of environment and training.}
\centering\footnotesize
\begin{tabular}{l|r}
\toprule[1pt]
    \textbf{Simulation Parameter} & \textbf{Value} \\ \midrule\midrule 
    UAV speed $(\nu)$& $45$\,km/h \\
    Target speed & $36$\,km/h \\
    URLLC range $(d^c)$ & $938$\,m \\ 
    Collision distance $(d^i)$ & $563$\,m \\
    Carrier frequency $(f^g_c)$ & $2$\,GHz \\
    Bandwidth $(W)$ & $20$\,MHz \\
    Payload size $(L_B)$ & $576$\,bits\\
    Transmit Power $(P_t)$ & $46$\,dBm \\
    Noise Power $(P_n)$ & $-99$\,dBm \\
    Time step per episode & $20$ \\ \toprule[1pt]
    \textbf{Training Parameter} & \textbf{Value} \\ \midrule\midrule
    Default number of nodes $(J_1)$ & $64$ \\
    Number of node in attention layers $(J_2)$ & $32$ \\
    epsilon-greedy constant $(\delta)$ & $0.3$ \\ 
    Annealing step & $1,000$ \\ 
    Training iterations & $5,000$ \\ 
    Batch size $(I)$ & $64$ \\
    Learning rate of GAXNet & $8\times10^{-4}$ \\
    Learning rate of QMIX & $1\times10^{-4}$ \\
    \bottomrule[1pt]
\end{tabular}

    \vspace{-5pt}
\end{table}

\begin{figure}[t!]
    \centering
    \includegraphics[width=0.55\columnwidth]{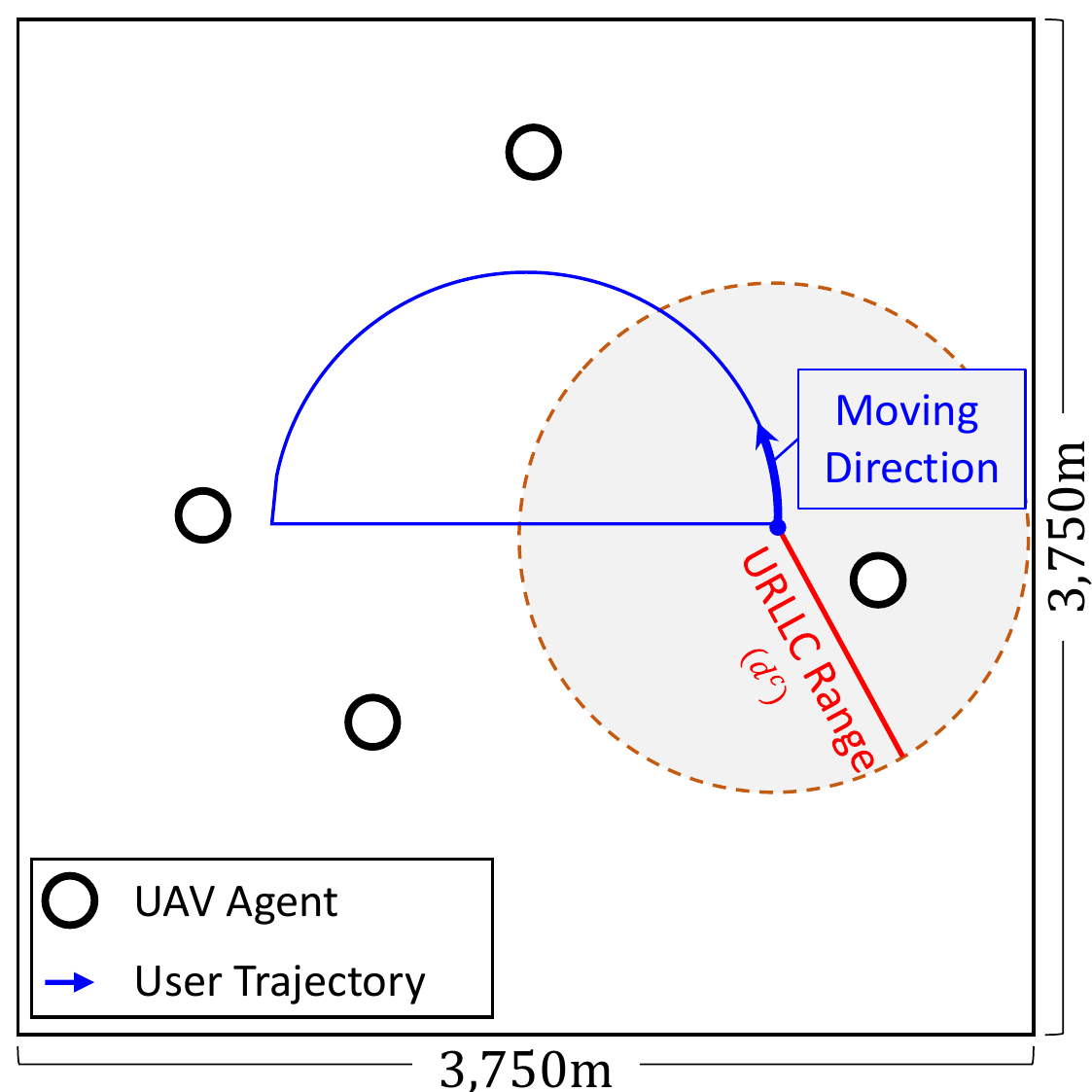}
    \caption{The simulation environment}
    \label{fig:environment}
\end{figure}

In order to verify the potential of the proposed approach, the environment is configured as shown in Fig.~\ref{fig:environment}.
The environment has a 2D $3,750$m $\times$ $3,750$m grid. $4$ UAV agents are located in the grid (i.e., $N=4$), a server is also located in the center of the grid, and users are also configured for moving around the server.
The experiment is devised by comparing the GAXNet with QMIX where the QMIX is the state of the art of CTDE.
The experiment was conducted using Python 3.6 and Pytorch. In addition, Table~\textsc{I} presented for specific experimental parameters. Moreover we set $\alpha=9.61$, $\beta=0.16$, $\eta_{LoS}=1$ [dB], and $\eta_{NLoS}=20$ [dB].
For fixed $L_B$, the transmission time is given as $T_\mathrm{max}$ and it can be assumed that $T_\mathrm{max}$ is less than the latency requirements. Actually, $T_\mathrm{max}$ can be calculated as 28.8 $\mu \text{s} $ given $L_B=576~\text{bits}$ and $W=20$MHz and it is within the 5G NR transmit latency \cite{WC_2018_latency}.
To meet the reliability requirements, $\varepsilon\leq \varepsilon_0$ needs to be satisfied.
By \eqref{error_rate} and the property of monotonic increasing function $f(\text{SNR})$, the required SNR is
\begin{align}\label{target_SNR}
    \text{SNR}\geq f^{-1}(Q_f^{-1}(\varepsilon_0)),
\end{align}
where $f^{-1}(\cdot)$ and $Q_f^{-1}(\cdot)$ is the inverse function of $f(\cdot)$ and $Q_f(\cdot)$, respectively.
Due to \eqref{SNR}, \eqref{target_SNR} can be rewritten as
\begin{align}\label{target_radius}
    \text{PL}(d,h)\leq -10\log_{10}\left(\frac{P_n}{P_t}f^{-1}\left(Q_f^{-1}(\varepsilon_0)\right)\right).
\end{align}

\begin{figure}[t!]
    \centering \small
\includegraphics[width=0.8\columnwidth]{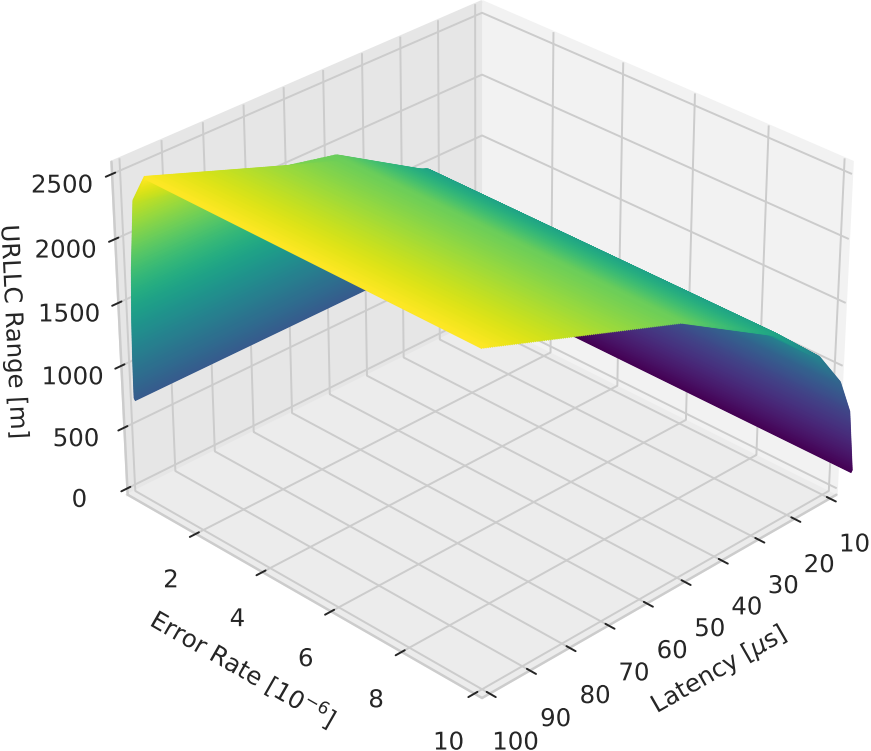} 
\caption{A2G network specification for URLLC: The relationship between three components; reward distance, error rate, and latency.}\label{fig:3Dplot}
\end{figure}

If the equality holds in \eqref{target_radius} when $d=d^*$, the reliability requirement is always satisfied with $d\leq d^*$ because the path loss in \eqref{PL} monotonically increases with $d$.
Therefore, we can obtain the URLLC range $d^c$ to satisfy the reliability requirements, $\varepsilon_0$, using \eqref{target_radius}.
Fig.~\ref{fig:3Dplot} shows the relationship between reward distance, required latency, and required error rate for URLLC via \eqref{target_SNR}--\eqref{target_radius}. For simulation, we adopt the required error rate as $10^{-7}$, the required latency as $39\mu\text{s}$, and the URLLC range as $938\text{m}$ for URLLC reliability.

\begin{figure}[t!]
    \centering
    \includegraphics[width=0.9\columnwidth]{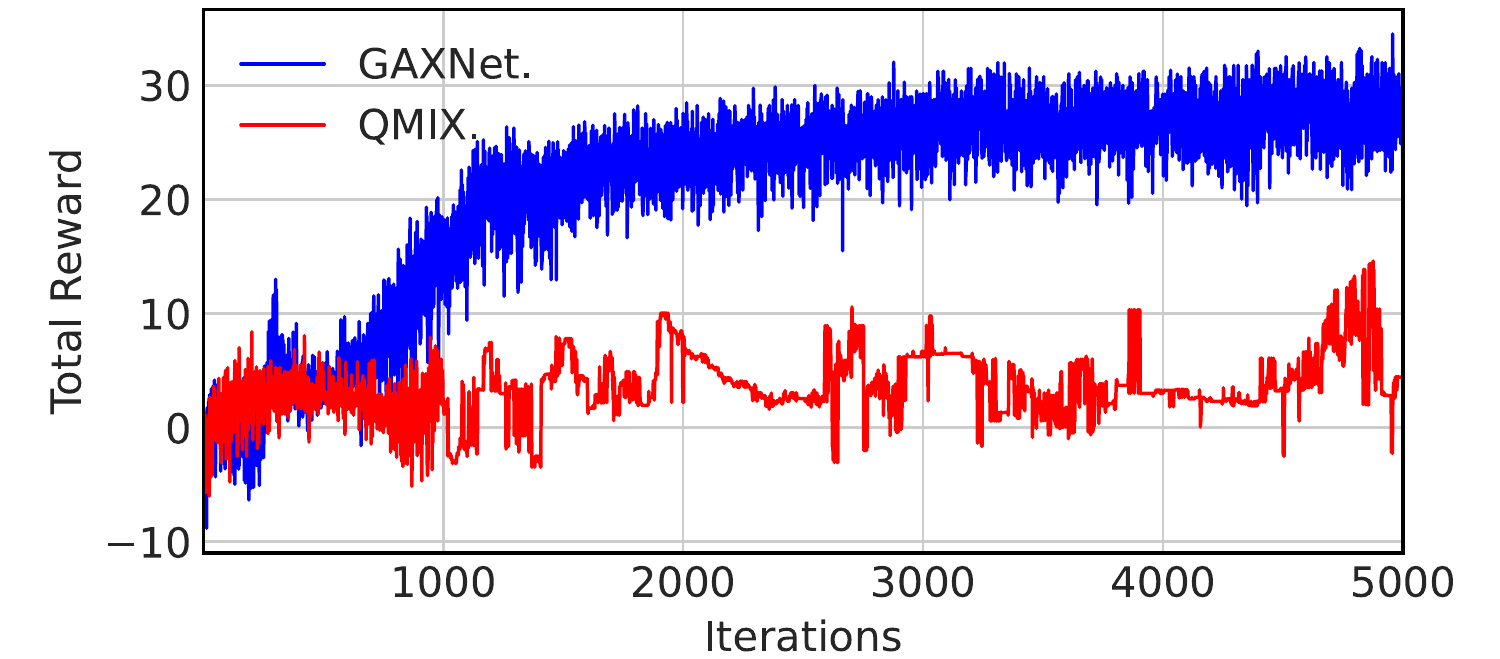}
    \caption{The learning curve of two algorithms.}
    \label{fig:reward}
\end{figure}

\begin{figure}[t!]
\centering \small
\begin{tabular}{cc}
       \includegraphics[width=0.45\linewidth]{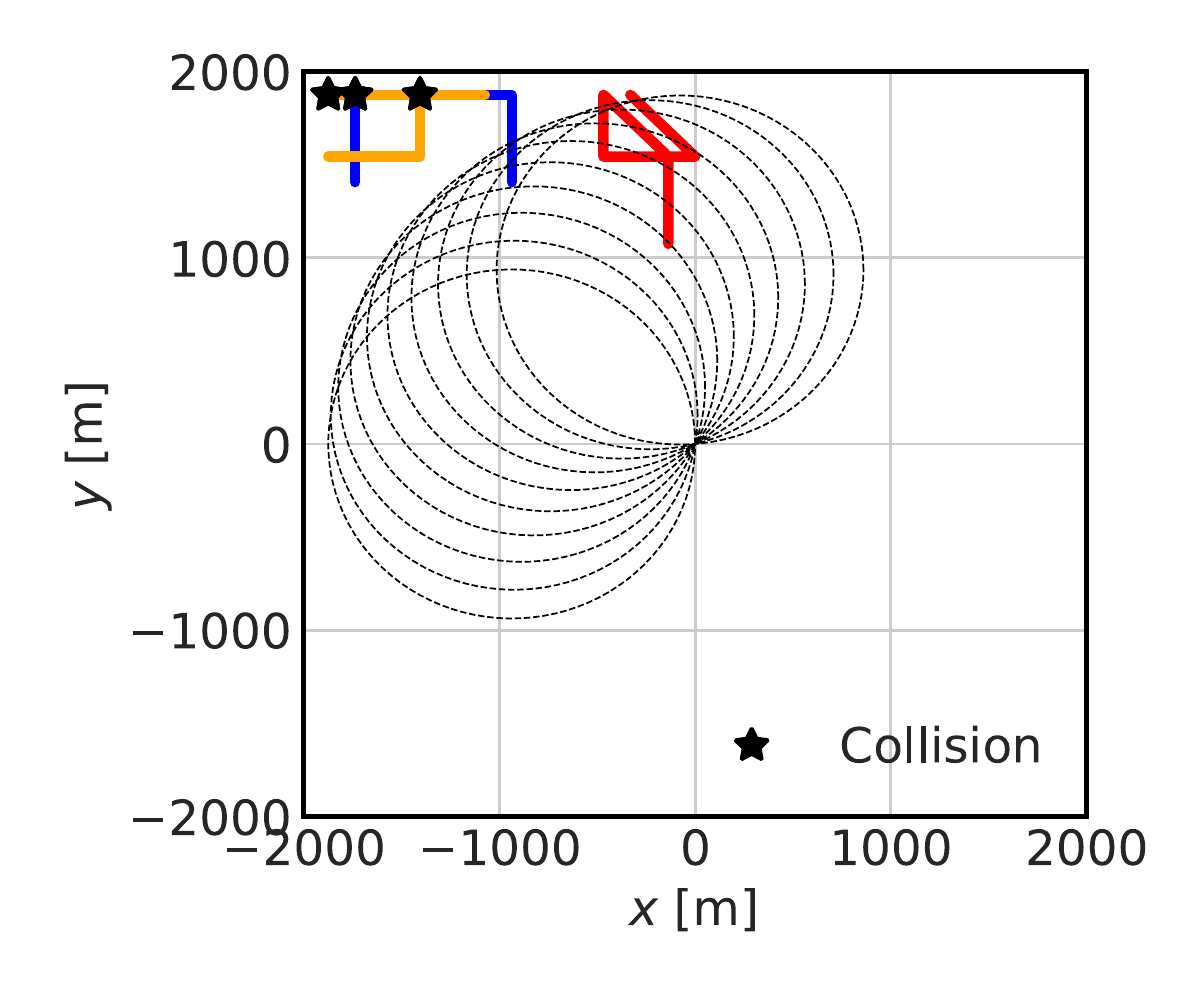}& \includegraphics[width=0.45\linewidth]{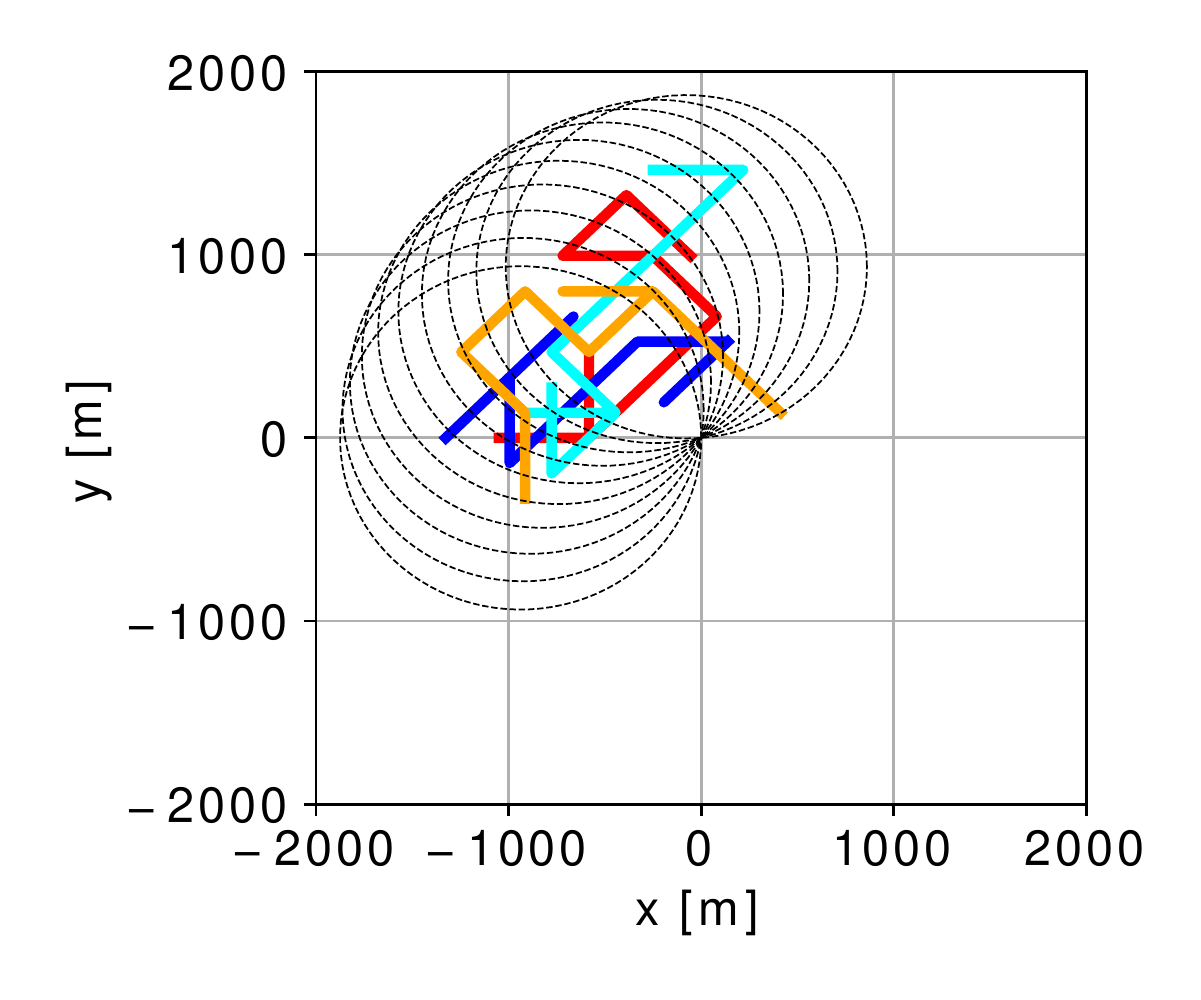}\\
    (a) QMIX. & (b) GAXNet.\\
\end{tabular}
    \caption{The trajectory of UAV agents in 10 time slots. The \textit{dotted circle}, \textit{star marker}, \textit{bold line} represent the target area, collision area, the trajectory of UAV agents, respectively.}
    \label{fig:traj}
\end{figure}

In reinforcement learning, we observe reward convergences, collisions, and target arrivals for two algorithm, i.e., GAXNet and QMIX-based algorithms. 
Fig.~\ref{fig:reward} shows the sum of total reward. As shown in Fig.~\ref{fig:reward}, the reward of GAXNet converges around to $26$ at $3,100$ iterations. However, the reward of QMIX fluctuates between $-3$ and $11$.  
Fig.~\ref{fig:traj} represents the visual trajectory dynamics of UAV agents. Fig.~\ref{fig:traj}(a) illustrates the UAV trajectory path planning with QMIX-based algorithm whereas Fig.~\ref{fig:traj}(b) visually presents the UAV trajectory dynamics with GAXNet. As shown in Fig.~\ref{fig:traj}(a), all agents do not trace the target, and there are collisions between agents. However, all agents using our purposed algorithm chase the target with out any collision as shown in Fig.~\ref{fig:traj}(b). In addition, collision between agents with GAXNet does not occur. 

\begin{figure}[t!]
\centering \small
\begin{tabular}{cc}
       \includegraphics[width=0.45\columnwidth]{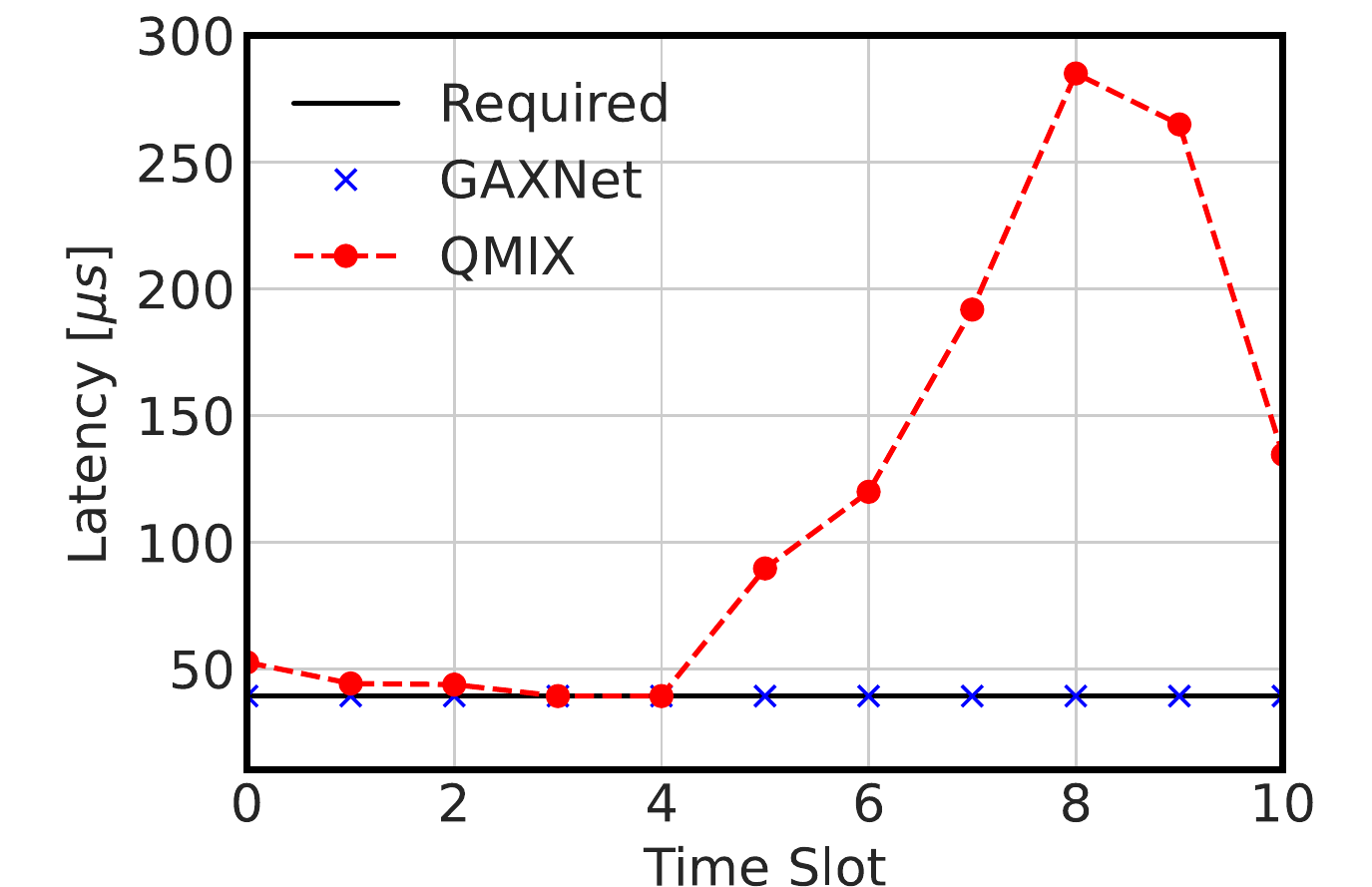}&
       \includegraphics[width=0.45\columnwidth]{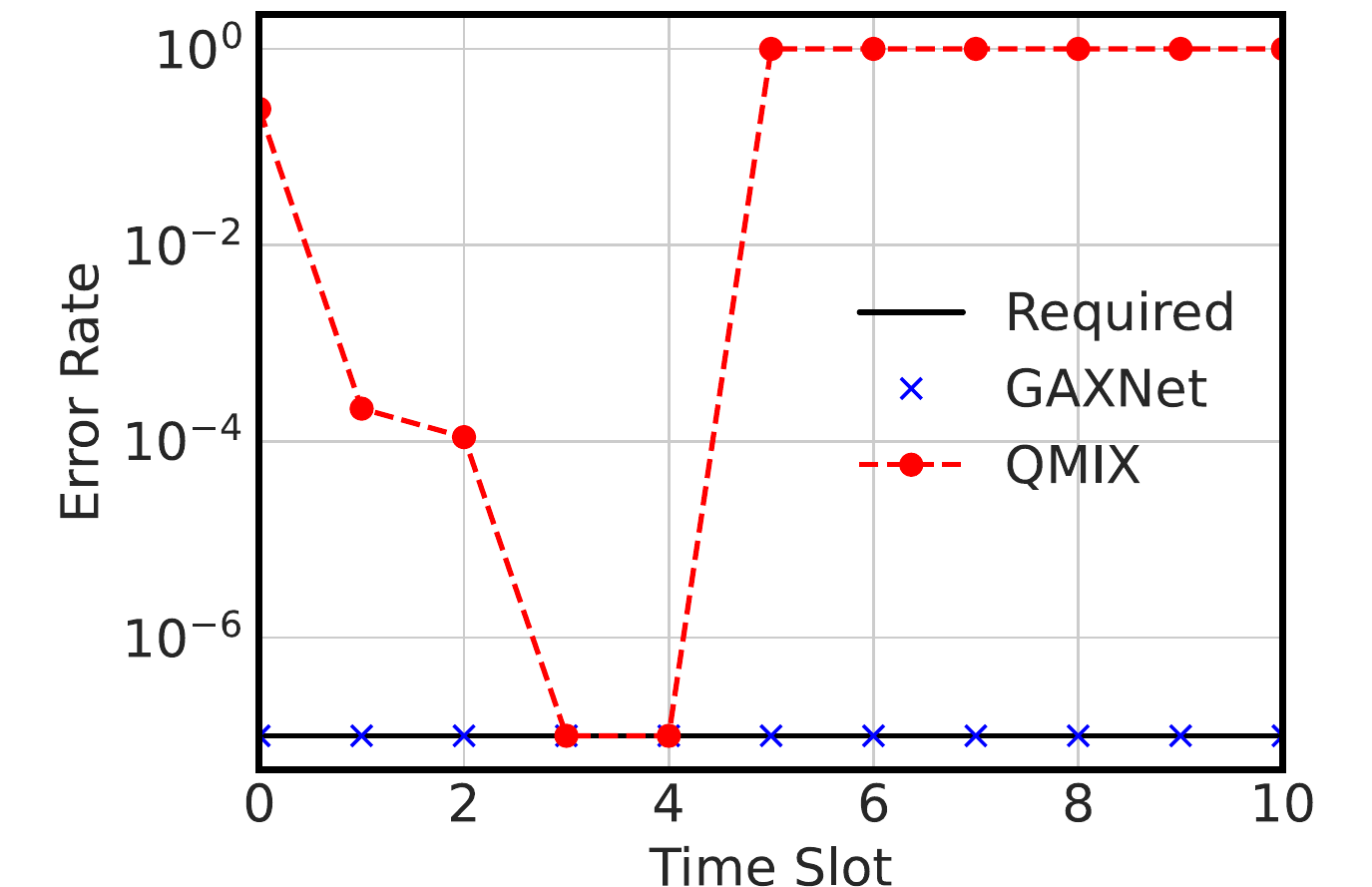}\\
       (a) Latency. & (b) Error rate. \\
\end{tabular}
    \caption{Comparison of two algorithms in respect to URLLC reliability in 10 time slots.}
    \label{fig:performance_urllc}
\end{figure}

In terms of URLLC, Fig.~\ref{fig:performance_urllc} shows the reliability of both latency and error rate. The GAXNet always satisfy the latency reliability $(39\mu\text{s})$ and error rate reliability $(10^{-7})$. 
However, QMIX only satisfy at the time slots [3; 4]. During the rest of time slots, the QMIX-based UAV agents communicate under high latencies and error rates. This implies that QMIX-based UAV agents may not be able to guarantee URLLC reliability.

Therefore GAXNet outperforms the baseline regarding the objectives of the system. 

\begin{figure}[t!]
    \centering \small
        \begin{tabular}{cc}
            \includegraphics[width=0.45\linewidth]{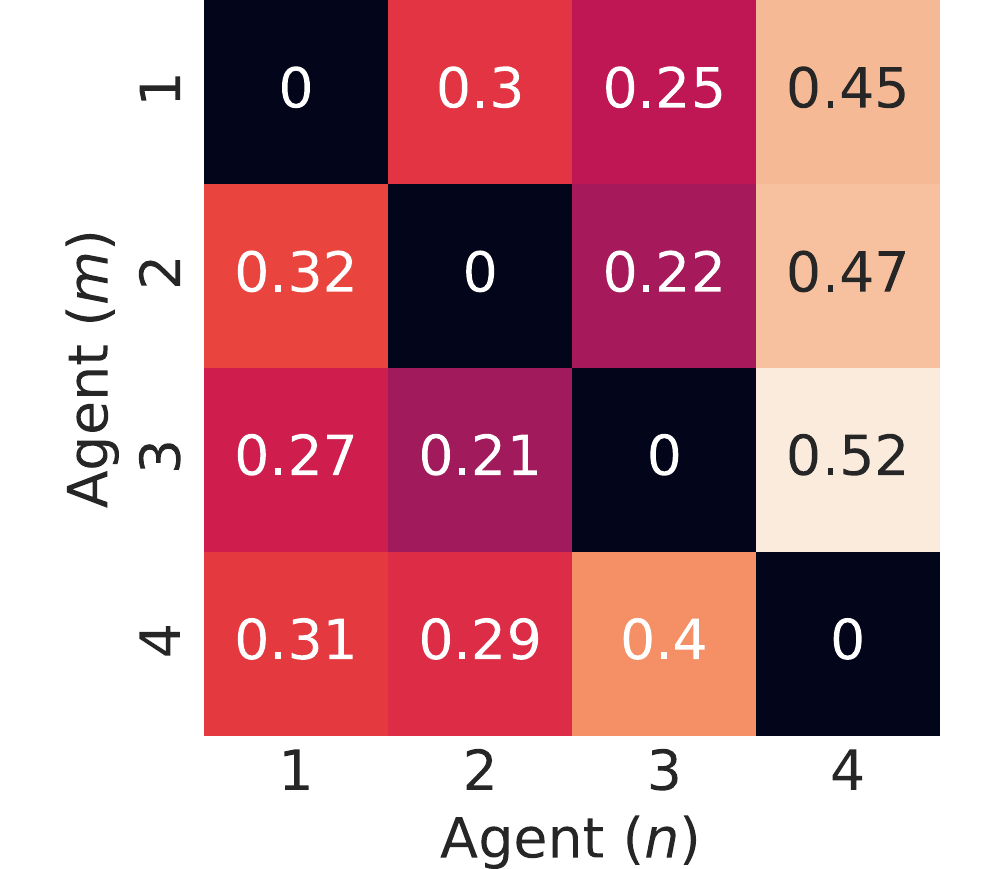}& \includegraphics[width=0.45\linewidth]{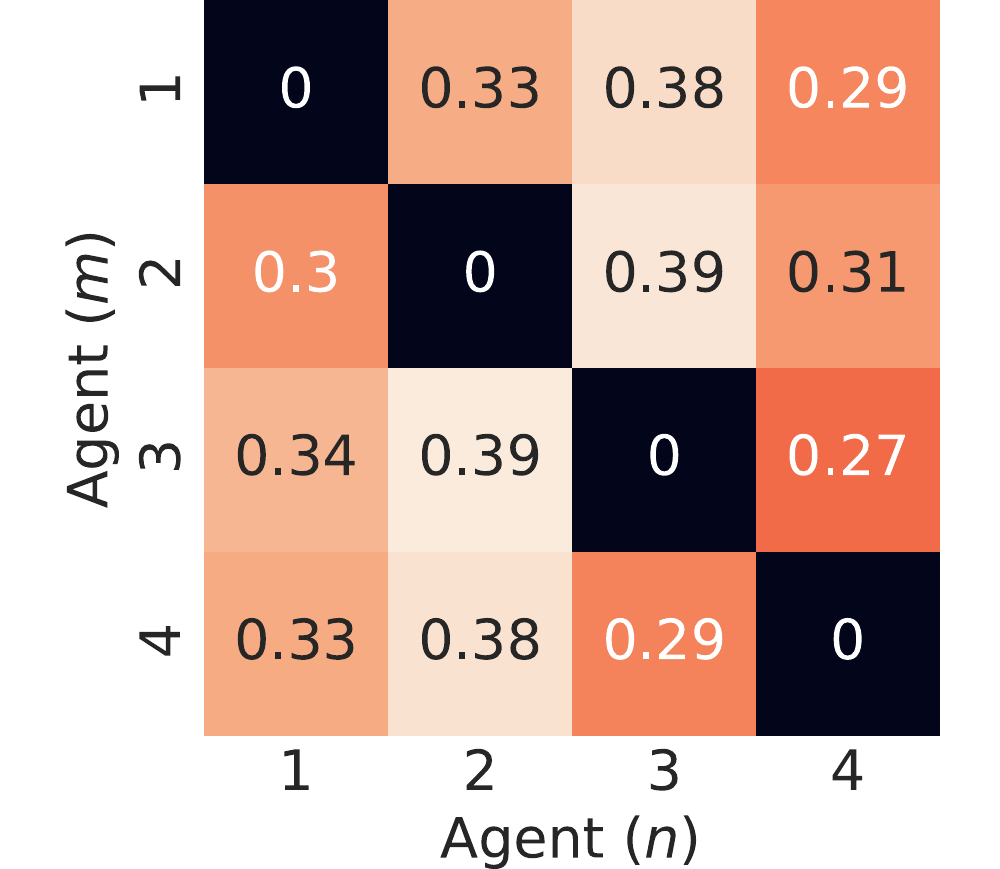}\\
            (a) GAXNet without exchange. & (b) GAXNet with exchange. \\
        \end{tabular}
        \caption{Semantic attention weights.}
        \label{fig:weight}
    \vspace{-15pt}
\end{figure}

We investigate the differences between untrained GAXNet without exchanging semantic weights, and trained GAXNet with exchanging. Fig.~\ref{fig:weight} represents the weighted adjacency matrix of $4$ agents. In the case of GAXNet without exchanging semantic weights, the agents which have the maximum difference between $\bar{w}^t_{n,m}$ and $w^{t-1}_{m,n}$ are $u^2$ and $u^4$. The maximum difference is $0.18$. Calculating the mean squared error (MSE) of all anti-diagonal elements (i.e., $\bar{w}^t_{n,m}$ and $w^{t-1}_{m,n}$), the result is $0.043$.
On the other hand, in the case of GAXNet with exchange, the maximum difference of two weights is $0.07$, and MSE is $0.014$. There is correlation of $o^{oth,n,m}$ with $o^{oth,m,n}$, because the relative position of $(u^n,u^m)$ has a relationship (\textit{e.g.}, $l^{n,m}_t = - l^{m,n}_t$) and the distance and connectivity information between $u^n$ and $u^m$ are mutually same (\textit{e.g.}, $d^{n,m}_t= d^{m,n}_t$ and $e^{n,m}_t = e^{m,n}_t$). For these reason, the attention weight between two agents (\textit{e.g.}, $w_{n,m}^t$ and $w_{m,n}^t$ should be similar. In other words, the weighted adjacency matrix as shown in Fig.~\ref{fig:weight} should be diagonal. As shown in Fig.~\ref{fig:weight}(a)/(b), GAXNet with exchange makes weighted adjacency matrix more symmetric than GAXNet without exchange. 
Because the SR encoder constructs symmetric matrix, we conclude that SR encoder is able to build semantic representation, successfully. 
\section{Conclusion}\label{sec:5}
In this paper, we developed a novel CTDE MADRL framework for UAV aided URLLC. The proposed solution, GAXNet, was shown to achieve lower latency with higher reliability compared to a state-of-the-art CTDE method, QMIX. To generalize and improve and extend GAXNet, incorporating realistic inter-UAV channels as well as considering a continuous UAV control action space could be interesting topics for future research.

\end{document}